\documentclass[prd,aps,eqsecnum,showpacs]{revtex4}
\usepackage{amssymb}
\usepackage{bm}
\usepackage[dvips]{graphicx}

\begin{document}

\title{Multi-brid inflation and non-Gaussianity}

\author{Misao {\scshape SASAKI}
}
\affiliation{
Yukawa Institute for Theoretical Physics, Kyoto University, 
Kyoto 606-8502, 
Japan
}

\vspace{5mm}

\date{\today}

\begin{abstract}
We consider a class of multi-component
hybrid inflation models whose evolution may be
analytically solved under the slow-roll approximation.
We call it multi-brid inflation (or $n$-brid inflation
where $n$ stands for the number of inflaton fields).
As an explicit example, we consider a 
two-brid inflation model, in which the inflaton potentials
are of exponential type and a waterfall field that
terminates inflation has the standard quartic potential
with two minima. Using the $\delta N$ formalism, we derive an expression
for the curvature perturbation valid to full nonlinear order.
Then we give an explicit expression for the curvature perturbation to
second order in the inflaton perturbation. We find that the final form
of the curvature perturbation depends crucially on
how the inflation ends. Using this expression, we present
closed analytical expressions for the spectrum of the curvature
perturbation ${\cal P}_{S}(k)$,
the spectral index $n_S$, the tensor to scalar ratio $r$,
and the non-Gaussian parameter $f_{NL}^{\rm local}$, in
terms of the model parameters.
We find that a wide range of the parameter space
$(n_S,\,r,\,f_{NL}^{\rm local})$ can be covered by
varying the model parameters within a physically reasonable range.
In particular, for plausible values of the model parameters,
we may have a large non-Gaussianity $f_{NL}^{\rm local}\sim 10$--$100$.
This is in sharp contrast to the case of 
single-field hybrid inflation in which these parameters
are tightly constrained.

\end{abstract}

\hfill YITP-08-35

\maketitle

\section{Introduction}
\label{sec:intro}
The standard, single-field slow-roll inflation
predicts the curvature perturbations which are
almost Gaussian to high accuracy~\cite{Maldacena:2002vr,Seery:2005wm}.
Thus the detection or non-detection of non-Gaussianity 
will have extremely important implications to theories 
of the early universe, and a variety of multi-component models 
that give detectable non-Gaussianities have been
proposed~\cite{NGmodels1,NGmodels2,Bernardeau:2002jf,
Dvali:2003em,Kofman:2003nx,Lyth:2005qk,Alabidi,Chambers:2007se}.
In many of these cases studied so far, however, it is not
quantitatively clear how the non-Gaussianity arises
and what determines its level.

In this paper, we study a class of multi-component
hybrid-type inflation models whose dynamics
can be analytically solved. Similar exactly soluble slow-roll
models were investigated in Ref.~\cite{Sasaki:2007ay}.
Here we extend the analysis given in Ref.~\cite{Sasaki:2007ay}
by including a coupling to a water-fall field which
terminates the inflationary stage. 
Then applying the $\delta N$
formalism~\cite{Sasaki:1995aw,Sasaki:1998ug,Wands:2000dp,Lyth:2004gb},
we compute the curvature perturbation analytically. 
Models of hybrid inflation similar to the one presented in this paper
were proposed and investigated by Bernardeau and Uzan~\cite{Bernardeau:2002jf}
and by Alabidi and Lyth~\cite{Alabidi}.

In the $\delta N$-formalism, the final amplitude of the curvature
perturbation on comoving hypersurfaces ${\cal R}_c$ 
(or equivalently on uniform total density hypersurfaces $\zeta$)
is given by $\delta N$, where $\delta N$ is the perturbation of the
number of $e$-folds between the initial flat time-slice
at horizon crossing during inflation and a comoving time-slice during the
final radiation-dominated stage by which time all the inflationary trajectories
are assumed to have converged to a unique one~\cite{Sasaki:1998ug,Lyth:2004gb}.

For the models we consider in this paper, assuming the scalar field
perturbations are Gaussian up to the time of horizon crossing,
the non-Gaussianity is local in the sense that the curvature
perturbation at each spatial point is expressed in terms of
a nonlinear function of Gaussian fields at the same spatial point.
In this case, the level of non-Gaussianity is conveniently characterized
by the quantity $f_{NL}^{\rm local}$~\cite{Komatsu:2001rj}.

This paper is organized as follows. In Section~\ref{sec:exact},
we first consider conditions for a slow-roll model to be
exactly soluble and derive a general formula for the curvature
perturbation. In Section~\ref{sec:model}, as a specific example,
we present a model of multi-field hybrid inflation, which we call
``{\it multi-brid inflation}''. Then focusing on the case of two-brid
inflation, we give various formulas explicitly.
In Section~\ref{sec:curvaturepert}, we compute analytically 
the spectrum of the curvature perturbation ${\cal P}_S(k)$, 
its spectral index $n_S$, the tensor-to-scalar ratio $r$,
and the non-Gaussian parameter $f_{NL}^{\rm local}$.
They are all expressed explicitly in terms of the model parameters.
We conclude the paper in Section~\ref{sec:conclusion}.
Some computational details are described in Appendix~\ref{sec:detail}.
We use the Planck units where $M_{pl}^{-2}=8\pi G=1$.

\section{Exact soluble class}
\label{sec:exact}

We consider an Einstein-scalar Lagrangian,
\begin{eqnarray}
L=\frac{1}{2}R
-\frac{1}{2}g^{\mu\nu}h_{ab}(\phi)\partial_\mu\phi^a\partial_\nu\phi^b
-V(\phi)\,,
\label{action}
\end{eqnarray}
where $R$ is the Ricci scalar, the Latin indices $a$, $b$ run from $1$ to $M$,
$h_{ab}$ is the field space metric, and $V$ is the potential. 
We assume the dynamics of inflation is described by this Lagrangian.
Later we introduce a water-fall field $\chi$ that terminates inflation.
For the moment, however, we concentrate our discussion on
the inflationary stage.

The Friedmann equation and the field equations are, respectively,
\begin{eqnarray}
&&3H^2=\frac{1}{2}h_{ab}\dot\phi^a\dot\phi^b+V(\phi)\,,
\cr
&&\ddot\phi^a+3H\dot\phi^a+h^{ab}\partial_bV=0\,,
\label{eom}
\end{eqnarray}
where $H=\dot a/a$ with $a$ being the cosmic scale factor,
and a dot denotes a derivative with respect to the cosmic
proper time $t$, $\dot{}=d/dt$.
The slow-roll equations of motion are obtained by
neglecting the kinetic term in the Friedmann equation
and the second time derivative in the field equation,
\begin{eqnarray}
&&3H^2=V(\phi)\,,
\cr
&&3H\dot\phi^a+h^{ab}\partial_bV=0\,.
\label{slowrolleom}
\end{eqnarray}
For later convenience, we change the time variable from
$t$ to the number of $e$-folds from the end of inflation
{\it backward\/} in time,
\begin{eqnarray}
dN=-Hdt\,.
\label{Ndef}
\end{eqnarray}
Then the slow-roll equations of motion give
\begin{eqnarray}
\frac{d\phi^a}{dN}=\frac{h^{ab}\partial_bV}{3H^2}
=\frac{h^{ab}\partial_bV}{V}\,.
\label{slowrollN}
\end{eqnarray}

Now we consider the case when the slow-roll equations of
 motion~(\ref{slowrollN}) can be exactly soluble.
They can be exactly solved if the 
right-hand side of Eq.~(\ref{slowrollN}) for each index
$a$ takes the form,
\begin{eqnarray}
\frac{h^{ab}\partial_bV}{V}
=\frac{f^a(\phi^a)}{F(\phi)}\,,
\label{solcond}
\end{eqnarray}
where $F$ represents an arbitrary function of $(\phi^1,\phi^2,\cdots,\phi^M)$,
and $f^a$ is a function of only $\phi^a$ for each $a$ ($a=1,2,\cdots,M$).
In this case, we have
\begin{eqnarray}
\frac{1}{f^a(\phi^a)}\frac{d\phi^a}{dN}=\frac{1}{F(\phi)}\,.
\end{eqnarray}
Introducing a new set of field coordinates $q^a$ by
\begin{eqnarray}
\ln q^a=\int\frac{d\phi^a}{f^a}\,,
\label{qdef}
\end{eqnarray}
the equations of motion become
\begin{eqnarray}
\frac{d\ln q^a}{dN}=\frac{1}{F}\,.
\label{qaeqn}
\end{eqnarray}
Here we introduce the radial and angular coordinates of 
the field space,
\begin{eqnarray}
q^a=q\,n^a\,;\quad \sum_a(n^a)^2=1\,.
\end{eqnarray}
Then it is straightforward to see that $n^a$ is conserved,
\begin{eqnarray}
\frac{dn^a}{dN}=0\,,
\label{dndN}
\end{eqnarray}
and Eq.~(\ref{qaeqn}) reduces to a single equation for $q$,
\begin{eqnarray}
\frac{d\ln q}{dN}=\frac{1}{F(q,n^a)}\,,
\label{qeq}
\end{eqnarray}
where $F$ is now regarded as a function of $q$ and $(n^1,n^2,\cdots,n^M)$.
\begin{center}
\begin{figure}
\includegraphics[width=12cm]{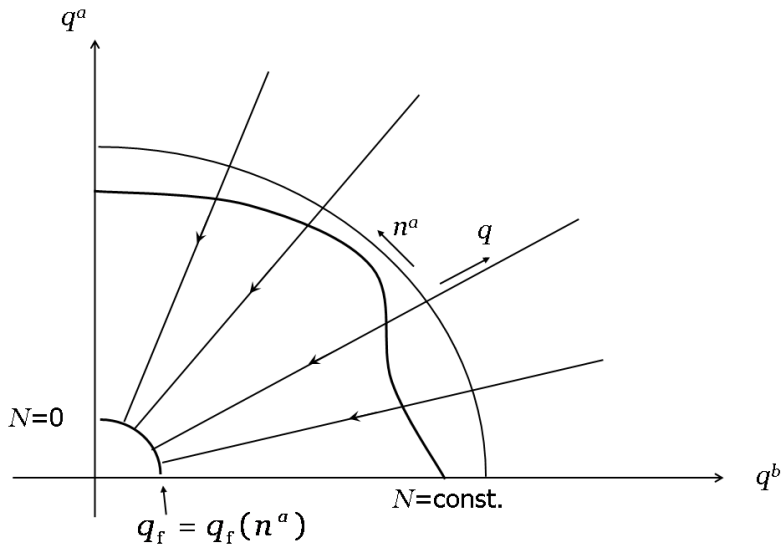}
\caption{A schematic diagram of classical trajectories 
in the field space with the coordinates $q^a$. 
The angular coordinates $n^a=q^a/q$ are conserved, and
hence all the trajectories are radial in these coordinates.
The curve indicated by $q=q_f$ denotes the surface
at which the inflation ends, which may depends on
$n^a$. The $e$-folding number $N$ is counted backward in time
from $q=q_f$.}
\label{fig:qspace}
\end{figure}
\end{center}

In fact, what we have shown is not restricted to slow-roll
inflation. Whenever a system approaches an attractor-like
asymptotic stage, the value of the field at one instant of time
completely determines the motion of the field at its subsequent time.
In such a case, we have $M$ first-order differential equations instead
of $M$ second-order differential equations. Then the exact solubility means
there are $M-1$ constants of integration. This is indeed the case
we have considered in the above (there are $M-1$ degrees of freedom in $n^a$).
A visualization of this situation is depicted in Fig.~\ref{fig:qspace}.

At this point, let us consider a couple of somewhat more specific examples.
The condition for exact solvability (\ref{solcond}) is satisfied 
if one can choose a set of field space coordinates such that
the field space metric takes the form,
\begin{eqnarray}
h^{ab}=\frac{{\rm diag.}
\left(h^1(\phi^1),h^2(\phi^2),\cdots,h^n(\phi^n)\right)}{H(\phi)}\,,
\end{eqnarray}
and the potential is ether of product type,
\label{prodpot}
\begin{eqnarray}
V=\prod_{a}V^a(\phi^a)\,,
\end{eqnarray}
in which case we have
\begin{eqnarray}
f^a=h^a\frac{d\ln V^a}{d\phi^a}\,,
\quad F=H\,.
\end{eqnarray}
or of separable type,
\begin{eqnarray}
V=\sum_aV^a(\phi^a)\,,
\label{seppot}
\end{eqnarray}
in which case we have
\begin{eqnarray}
f^a=h^a\frac{dV^a}{d\phi^a}\,,
\quad F=HV\,.
\end{eqnarray}
The separable case was first investigated 
in detail by Starobinsky~\cite{Starobinsky:1986fxa}.

Now, going back to the general case, Eq.~(\ref{dndN})
implies the trajectories are always radial in the
field space spanned by $q^a$, and Eq.~(\ref{qeq})
 can be easily solved for $N$ to give
\begin{eqnarray}
N(q,n^a)=\int_{q_f}^qF(q',n^a)\,d\ln q'\,,
\label{efoldsol}
\end{eqnarray}
where $q_f$ is the value of $q$ at the end of inflation.
Note that $q_f$ generally depends on $n^a$.
Thus, applying the nonlinear $\delta N$ formula,
we immediately obtain a fully nonlinear expression
for the conserved comoving curvature perturbation,
\begin{eqnarray}
\delta N=N(q+\delta q,n^a+\delta n^a)-N(q,n^a)\,.
\label{NLdeltaN}
\end{eqnarray}

It is instructive to consider the linear limit of
the above $\delta N$. Denoting it by $\delta_LN$,
we find
\begin{eqnarray}
\delta_LN=F(q,n^a)\frac{\delta q}{q}
+\int_{q_f}^q\frac{\partial F}{\partial n^a}\frac{dq'}{q'}\,\delta n^a
-\frac{F}{q_f}\frac{\partial q_f}{\partial n^a}\,\delta n^a\,,
\label{dNlin}
\end{eqnarray}
where the values of $(q,n^a)$ are those at horizon crossing during
inflation and $(\delta q,\delta n^a)$ are the fluctuations 
evaluated on the flat hypersurface at that epoch.
It may be noted that $\delta q^a$ are given by the 
fluctuations of the original field variables $\delta\phi^a$ as
\begin{eqnarray}
\delta\ln q^a=\frac{\delta\phi^a}{f^a(\phi^a)}\,.
\end{eqnarray}
 The formula (\ref{dNlin}) consists of three terms.
The first term may be interpreted as the one due to adiabatic perturbations,
the second to entropy perturbations during
 inflation~\cite{Polarski:1994rz,Mukhanov:1997fw,Sasaki:1998ug},
and the third to entropy perturbations at the end of 
inflation~\cite{Dvali:2003em,Kofman:2003nx,Lyth:2005qk}.
It may be worthwhile to mention that although these distinctions are
meaningful and useful at linear order (or perhaps perturbatively),
they are not so at full nonlinear order.

\begin{center}
\begin{figure}
\includegraphics[width=12cm]{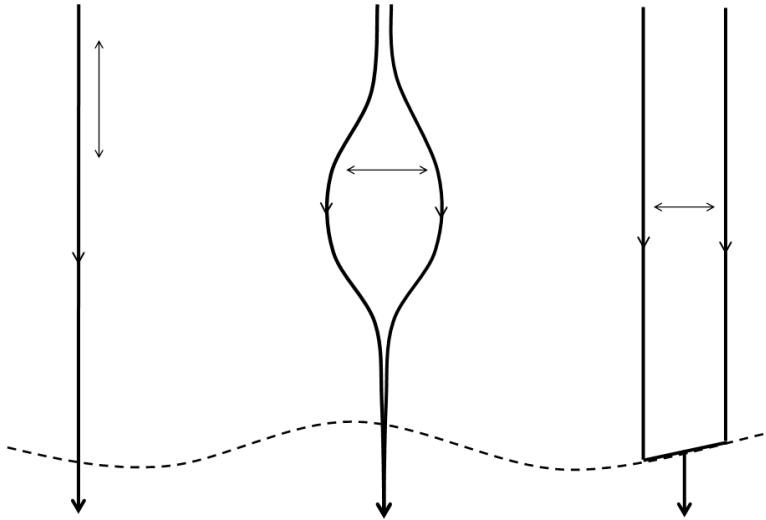}
\caption{Schematic graphs describing the three terms in Eq.~(\ref{dNlin}).
The thick lines represent three different kinds of orbits in the field space,
and the thin lines with arrows on both ends represent the field
fluctuations. The wavy dashed line represents the end of inflation.
The one on the left corresponds to the first term,
the one in the center to the second term, and the one on the right to
the third term. See text for more details.
}
\label{fig:orbits}
\end{figure}
\end{center}

Figure~\ref{fig:orbits} is a schematic graph to explain the meaning of these
three terms. 
The thick lines represent three different kinds of trajectories
in the field space,
and the thin lines with arrows on both ends represent the field
fluctuations. The wavy dashed line represents the surface at which
inflation ends. After inflation, each kind of orbit converges to
a unique trajectory.

The one on the left, where the fluctuations are parallel to
the orbits, corresponds to the first term in Eq.~(\ref{dNlin}). 
In this case, the fluctuations in the initial condition directly
gives $\delta N$. This is identical to the conventional adiabatic
curvature perturbation in the single field case.
The one in the center corresponds to the second term.
The fluctuations are orthogonal to the orbits, so they are entropy
(isocurvature) perturbations during inflation. However, by the time
the two adjacent orbits converge to a unique orbit, one finds the
number of $e$-folds depends on which route the universe has taken. 
This gives rise to $\delta N$ before the end of inflation.
The one on the right corresponds to the third term. The entropy perturbations
are not converted to $\delta N$ until the end of inflation. However, 
the surface that determines the end of inflation may not be orthogonal
to the orbits. This gives rise to $\delta N$ in the end.

\section{multi-brid inflation}
\label{sec:model}

We consider a multi-component inflaton field with the potential,
\begin{eqnarray}
V(\bm{\phi})=V_0\exp\left[\sum_{A=1}^Mm_A\phi_A\right]\,.
\label{potential}
\end{eqnarray}
Without loss of generality, we assume $m_A\phi_A>0$ for all $A=1,2,\cdots,M$.
The slow-roll equations of motion are
\begin{eqnarray}
\frac{d\phi_A}{dN}
=\frac{1}{V}\frac{\partial V}{\partial\phi_A}=m_A\,.
\label{mbrideom}
\end{eqnarray}
Note that the effective mass-squared $M_A^2$ for each
$\phi_A$ is given by
\begin{eqnarray}
M_A^2=\frac{\partial^2 V}{\partial\phi_A^2}
=m_A^2V=3\,m_A^2H^2\,.
\label{effM2}
\end{eqnarray}
Thus the slow-roll condition is satisfied for $m_A^2\ll1$.
Introducing a new variable $q_A$ by
\begin{eqnarray}
q_A=\exp[\phi_A/m_A]=q\,n_A\,;
\quad \sum_A(n_A)^2=1\,,
\end{eqnarray}
the slow-roll equations become
\begin{eqnarray}
\frac{d\ln q}{dN}=1\,,\quad \frac{dn_A}{dN}=0\,.
\end{eqnarray}
These are immediately integrated to give
\begin{eqnarray}
N=\ln q-\ln q_f=\frac{1}{2}
\ln\left[\sum_Ae^{2\phi_A/m_A}\right]
-\frac{1}{2}\ln\left[\sum_Ae^{2\phi_{A,f}/m_A}\right]\,.
\label{Nmultisol}
\end{eqnarray}

Now, similar to the conventional single-field hybrid
inflation~\cite{Linde:1993cn}, we assume that inflation ends at 
\begin{eqnarray}
\sum_Ag_A^2\phi_A^2=\sigma^2\,,
\label{endcond}
\end{eqnarray}
and the universe is thermalized instantaneously.
This may be realized by introducing a water-fall field
$\chi$ that terminates inflation. Specifically, we promote
$V_0$ in Eq.~(\ref{potential}) to a function of $\chi$ 
(and of $\phi_A$) as
\begin{eqnarray}
V_0=\frac{1}{2}\sum_Ag_A^2\phi_A^2\chi^2+\frac{\lambda}{4}
\left(\chi^2-\frac{\sigma^2}{\lambda}\right)^2\,.
\label{waterfallpot}
\end{eqnarray}
Then we see that the field $\chi$ is stable at the origin
for $\sum_Ag_A^2\phi_A^2>\sigma^2$, but becomes unstable
for $\sum_Ag_A^2\phi_A^2<\sigma^2$ which brings inflation
to an abrupt end.

Now we restrict our discussion to the case of two inflaton
fields, that is, two-brid inflation. Then it is convenient
to parametrize $n_A$ ($A=1,2$) as
\begin{eqnarray}
n_1=\cos\theta\,,\quad n_2=\sin\theta\,.
\label{thetadef}
\end{eqnarray}
Similarly, we parametrize the scalar field at the end of inflation as
\begin{eqnarray}
\phi_{1,f}=\frac{\sigma}{g_1}\cos\gamma\,,
\quad
\phi_{2,f}=\frac{\sigma}{g_2}\sin\gamma\,.
\label{gammadef}
\end{eqnarray}
Note that since $m_A\phi_A>0$ ($A=1,2$), both
$m_1g_1\cos\gamma$ and $m_2g_2\sin\gamma$ are positive.
Then, since $\theta$ is a constant of motion, we have
\begin{eqnarray}
\ln\left[\frac{q_1}{q_2}\right]=\frac{\phi_1}{m_1}-\frac{\phi_2}{m_2}
=\frac{\sigma\cos\gamma}{g_1m_1}-\frac{\sigma\sin\gamma}{g_2m_2}\,.
\label{constofmotion}
\end{eqnarray}
This equation may be solved for $\gamma$ in terms of $\phi_1$ and $\phi_2$.
Hence we may regard $\gamma$ as a function of $(\phi_1,\phi_2)$,
$\gamma=\gamma(\phi_1,\phi_2)$.
Therefore, the number of $e$-folds (\ref{Nmultisol}) may now
be regarded as a function of $(\phi_1,\phi_2)$,
\begin{eqnarray}
N=N(\phi_1,\phi_2)
=\frac{1}{2}\ln\left[\frac{\displaystyle
e^{2\phi_1/m_1}+e^{2\phi_2/m_2}}
{\displaystyle
e^{2\sigma\cos\gamma/(g_1m_1)}
+e^{2\sigma\sin\gamma/(g_2m_2)}}
\right].
\label{N2brid}
\end{eqnarray}
It is then straightforward to obtain $\delta N$ to full nonlinear order.
It is given simply by
\begin{eqnarray}
\delta N=N(\phi_1+\delta\phi_1,\phi_2+\delta\phi_2)-N(\phi_1,\phi_2)\,.
\label{nldeltaN}
\end{eqnarray}
Here it may be useful to clarify the origin of $\delta N$ in the present
model. Going back to the general expression for the linear
curvature perturbation (\ref{dNlin}), and noting that
we have $F=1$ in the equation of motion~(\ref{qeq}) in the present model,
we see that there is no contribution from the second term which is 
due to entropy perturbations during inflation, Thus the linear curvature
perturbation is due to the initial adiabatic perturbation at
the time of horizon crossing and to the final entropy perturbation
at the end of inflation. Explicit computations of these contributions to
$\delta N$ are given in Appendix~\ref{sec:details}.

Before closing this section, we note that there is a correction to be
added to the above formula in the rigorous sense.
It comes from the fact that the surface where the inflation ends
in the field space, Eq.~(\ref{endcond}), is not equal to the surface
of constant energy density. This means that the hot Friedmann stage
of the universe starts at slightly different temperatures
for different values of $(\phi_{1,f},\phi_{2,f})$ at which inflation
was terminated. This can be taken care of by introducing a correction
to the number of $e$-folds given by Eq.~(\ref{N2brid}).
Namely, denoting the potential energy at the end of inflation by $V_f$,
we add a correction term $N_c$ to Eq.~(\ref{N2brid}) as
\begin{eqnarray}
N=\frac{1}{2}\ln\left[\frac{\displaystyle
e^{2\phi_1/m_1}+e^{2\phi_2/m_2}}
{\displaystyle
e^{2\sigma\cos\gamma/(g_1m_1)}
+e^{2\sigma\sin\gamma/(g_2m_2)}}
\right]+N_c(\gamma)\,,
\end{eqnarray}
where 
\begin{eqnarray}
N_c=\frac{1}{4}\ln\left[\frac{V_f}{V_0}\right]
=\frac{\sigma}{4}\left(\frac{m_1}{g_1}\cos\gamma
+\frac{m_2}{g_2}\sin\gamma\right)\,.
\label{Ncorrect}
\end{eqnarray}
In the above we have assumed that the universe has become radiation-dominated
right after inflation.
It is then easy to see that this correction term is negligible for
sufficiently small $m_1$ and $m_2$. To be more specific,
if we compare the $\gamma$-dependent terms in the original $N$
to those in $N_c$, the dependence is apparently strong in the former
for small $m_1$ and $m_2$ irrespective of the values of $\sigma$,
$g_1$ and $g_2$. Explicit computations of $\delta N$ 
and $\delta N_c$ to second order in the field fluctuations are 
carried out in Appendix~\ref{sec:detail}.

\section{Curvature perturbation in two-brid inflation}
\label{sec:curvaturepert}

Let us evaluate the curvature perturbation explicitly. 
The quantities of our interest are the curvature perturbation
spectrum ${\cal P}_{S}$, the spectral index $n_S$, tensor-to-scalar
ratio $r$, and the non-Gaussianity $f_{NL}^{\rm local}$. 
To evaluate these quantities, we expand the $\delta N$ formula (\ref{nldeltaN})
to second order in $\delta\phi$ for $N(\phi)$ given in (\ref{N2brid}).
Note that $\delta\gamma$ must be expressed in terms of $\delta\phi$
with second order accuracy, using Eq.~(\ref{constofmotion}).
Details are given in Appendix~\ref{sec:detail}.
The result is
\begin{eqnarray}
\delta N
=\frac{\delta\phi_1\,g_1\cos\gamma+\delta\phi_2\,g_2\sin\gamma}
{m_1g_1\cos\gamma+m_2g_2\sin\gamma}
+\frac{g_1^2g_2^2}{2\sigma}
\frac{(m_2\delta\phi_1-m_1\delta\phi_2)^2}
{(m_1g_1\cos\gamma+m_2g_2\sin\gamma)^3}+\cdots\,.
\label{dN2nd}
\end{eqnarray}

We assume that the scalar field fluctuations $\delta\phi_1$ and $\delta\phi_2$
are Gaussian, with the dispersion,
\begin{eqnarray}
\left\langle\delta\phi_A\delta\phi_B\right\rangle_k
=\left.\left(\frac{H}{2\pi}\right)^2\right|_{t_k}\delta_{AB}\,,
\label{Gaussdphi}
\end{eqnarray}
where $t_k$ is the horizon-crossing time of the
comoving wavenumber $k$, $k=Ha$.
Then the curvature perturbation spectrum is given by
\begin{eqnarray}
{\cal P}_{S}(k)\equiv
\frac{4\pi k^3}{(2\pi)^3}P_{\cal R}(k)
=\frac{g_1^2\cos^2\gamma+g_2^2\sin^2\gamma}
{(m_1g_1\cos\gamma+m_2g_2\sin\gamma)^2}
\left.\left(\frac{H}{2\pi}\right)^2\right|_{t_k}\,.
\label{PcalR}
\end{eqnarray}
Using the fact that $3H^2=V$, the spectral index is easily calculated as
\begin{eqnarray}
n_S\equiv1+\frac{d\ln{\cal P}_{\cal R}(k)}{d\ln k}
=1-(m_1^2+m_2^2)\,.
\label{nS}
\end{eqnarray}
Thus our model predicts $n_S<1$.
As for the tensor-to-scalar ratio, it is given by
\begin{eqnarray}
r\equiv\frac{{\cal P}_T}{{\cal P}_S}
=8\frac{(m_1g_1\cos\gamma+m_2g_2\sin\gamma)^2}
{g_1^2\cos^2\gamma+g_2^2\sin^2\gamma}\,.
\label{TtoS}
\end{eqnarray}

As we see from Eqs.~(\ref{PcalR}), (\ref{nS}) and (\ref{TtoS}),
our model seems to have a sufficient number of parameters
that can be tuned to give the values of $n_S$ and $r$ 
which are consistent with observations. 
This is in contrast to a single-field hybrid inflation model
with $V=V_0\,e^{m\phi}$, for which the number of $e$-folds is given by
\begin{eqnarray}
N_{\rm single}=\frac{\phi-\phi_f}{m}\,,
\end{eqnarray}
which gives
\begin{eqnarray}
n_S=1-m^2\,,\quad r=8m^2
\quad(\mbox{single-field case})\,,
\end{eqnarray}
and, to first order in the slow-roll approximation,
the non-Gaussianity is exactly zero. Comparing
this single-field case with our two-field case, it is amazing that
adding only one extra component to the inflaton field results in
the huge variety of the output parameters.

Now we evaluate the non-Gaussianity in our model. For convenience,
we introduce the linear curvature perturbation ${\cal R}_L$ 
and the linear entropy perturbation $S$,
\begin{eqnarray}
{\cal R}_L=\frac{\delta\phi_1\,g_1\cos\gamma+\delta\phi_2\,g_2\sin\gamma}
{m_1g_1\cos\gamma+m_2g_2\sin\gamma}\,,
\quad
S=\frac{\delta\phi_1\,g_2\sin\gamma-\delta\phi_2\,g_1\cos\gamma}
{m_2g_1\cos\gamma-m_1g_2\sin\gamma}\,.
\label{NandS}
\end{eqnarray}
For the Gaussian fluctuations $\delta\phi_A$ given by Eq.~(\ref{Gaussdphi}),
we see that they are orthogonal to each other:
\begin{eqnarray}
\left\langle{\cal R}_L\cdot S\right\rangle=0\,.
\end{eqnarray}
In terms of ${\cal R}_L$ and $S$, the nonlinear $\delta N$ in Eq.~(\ref{dN2nd})
is expressed as
\begin{eqnarray}
\delta N={\cal R}_L+\frac{3}{5}f_{NL}^{\rm local}({\cal R}_L+S)^2\,,
\label{dNfin}
\end{eqnarray}
where
\begin{eqnarray}
f_{NL}^{\rm local}
=\frac{5g_1^2g_2^2}{6\sigma(g_1^2\cos^2\gamma+g_2^2\sin^2\gamma)^2}
\frac{(m_2g_1\cos\gamma-m_1g_2\sin\gamma)^2}
{m_1g_1\cos\gamma+m_2g_2\sin\gamma}\,.
\label{fNL}
\end{eqnarray}
Since both $m_1g_1\cos\gamma$ and $m_2g_2\sin\gamma$ are positive,
we thus see that in this model $f_{NL}^{\rm local}$ is always non-negative.
In addition, we note that this non-Gaussianity comes essentially from
the end of inflation, as seen from its explicit dependence on the 
parameters that define the end of inflation. This fact may be explicitly
seen by separating $\delta N$ into two parts; one from the perturbation
during inflation $\delta N_*$ and the other from the end of
inflation $\delta N_e$. This is done in Appendix~\ref{sec:detail}.

We also see that the linear entropy perturbation $S$ induces a curvature
perturbation at second order.  It may be worthwhile to note that if one
finds a way to extract out observationally the second order term, it may
be possible to detect this second order effect of the entropy perturbation.
In this connection, we mention recent work by Enqvist
et al.~\cite{Enqvist:2004ey}
and by Chambers and Rajantie~\cite{Chambers:2007se}, in which they discussed
a mechanism of generating non-Gaussianity at the end of inflation through
parametric resonance when there exist two light scalar fields.
Their mechanism corresponds effectively to the case where the second order
term in (\ref{dNfin}) is dominated by $S^2$. In this case, the curvature
perturbation may be approximately expressed as
\begin{eqnarray}
\delta N={\cal R}_L+f_S\,S^2\,.
\end{eqnarray}
Chambers and Rajantie~\cite{Chambers:2007se}
 define a non-Gaussian parameter essentially by the ratio,
\begin{eqnarray}
f_{NL}^{\rm CR}\sim\frac{\left\langle\delta N^3\right\rangle}
{\left\langle\delta N^2\right\rangle^2}\,.
\end{eqnarray}
Then, under the assumption that ${\cal R}_L$ dominates over $f_SS^2$
in $\delta N$, that is,
$f_S^2\left\langle S^2\right\rangle^2\ll\left\langle{\cal R}_L^2\right\rangle$,
one finds
\begin{eqnarray}
f_{NL}^{\rm CR}\sim f_S^3\,\frac{\left\langle S^2\right\rangle^3}
{\left\langle{\cal R}_L^2\right\rangle^2}\,.
\end{eqnarray}
In our case, since $f_S\sim f_{NL}^{\rm local}$ and 
$\left\langle S^2\right\rangle \sim \left\langle{\cal R}_L^2\right\rangle$,
our $f_{NL}^{\rm CR}$ would be small
$\sim (f_{NL}^{\rm local})^3\left\langle S^2\right\rangle$
 even for $f_{NL}^{\rm local}\sim100$.

To examine the viability of our model, let us compare
our predictions with observations.
Analysis of the WMAP 5-year data~\cite{Dunkley:2008ie,Komatsu:2008hk} gives
the spectral index of the value,
\begin{eqnarray}
n_S=0.96~\lower1ex\hbox{$\stackrel{\displaystyle+0.014}{-0.015}$}~,
\end{eqnarray}
and the upper limit on the tensor-to-scalar ratio,
\begin{eqnarray}
&&r<0.43\quad (95\%~\mbox{CL: WMAP alone})\,,
\nonumber\\
&&r<0.20\quad (95\%~\mbox{CL: WMAP$+$BAO$+$SN})\,.
\end{eqnarray}
As for the non-Gaussian parameter $f_{NL}^{\rm local}$,
the situation is somewhat controversial.
Yadav and Wandelt~\cite{Yadav:2007yy} finds non-vanishing non-Gaussianity
in the 3-year WMAP data~\cite{WMAP3y} as
\begin{eqnarray}
27 < f_{NL}^{\rm local} < 147\quad (95\%~\mbox{CL})\,,
\end{eqnarray}
while Komatsu et al. finds only a bound 
in the WMAP 5-year data~\cite{Komatsu:2008hk},
\begin{eqnarray}
-9<f_{NL}^{\rm local}<111\quad(95\%~\mbox{CL}) \,.
\end{eqnarray}
In any case, these results indicate that 
a positive value of $f_{NL}^{\rm local}$
is more favored over a negative value, if non-zero.
This is perfectly consistent with our result, Eq.~(\ref{fNL}).

To show that these observational data can be indeed 
reproduced in our model, let us consider a specific set
of the model parameters. We set
\begin{eqnarray}
m_1^2\sim0.005\,,\quad m_2^2\sim0.035\,,
\end{eqnarray}
and assume that the inflationary trajectory satisfies
\begin{eqnarray}
m_1\cos\gamma\gg m_2\sin\gamma\,,
\end{eqnarray}
that is, $\gamma\ll1$. We also assume
that the coupling constants $g_1$ and $g_2$ are
of the same order of magnitude.
In this case, $n_S$ and $r$ depend only on $m_1$ and $m_2$,
\begin{eqnarray}
n_S=1-(m_1^2+m_2^2)\sim0.96\,,\quad
r\approx 8m_1^2\sim 0.04\,.
\end{eqnarray}
On the other hand, the non-Gaussian parameter $f_{NL}^{\rm local}$
depends on the other model parameters.
For simplicity, let us assume
\begin{eqnarray}
g_1^2=g_2^2\equiv g^2\,.
\end{eqnarray}
Then the curvature perturbation spectrum is
approximately given by
\begin{eqnarray}
{\cal P}_S\approx\frac{1}{m_1^2}\left(\frac{H}{2\pi}\right)^2
\sim2.5\times10^{-9}\,,
\end{eqnarray}
where the second equality is from the WMAP normalization~\cite{Liddle:2006ev}.
Since we have fixed $m_1^2$ already, this normalization determines $H^2$,
and hence $\sigma^4$ as a function of $\lambda$,
\begin{eqnarray}
H^2\sim 40\, m_1^2\,{\cal P}_S\sim 5\times10^{-10}\,,
\quad
\sigma^2=(12\lambda H^2)^{1/2}\sim \lambda^{1/2}\times10^{-4}\,.
\label{HandL}
\end{eqnarray}
This gives
\begin{eqnarray}
f_{NL}^{\rm local}\approx\frac{5\,g\,m_2^2}{6\,m_1\sigma}
\sim40\,\frac{g}{\lambda^{1/4}}\,.
\label{fNLvalue}
\end{eqnarray}
Thus, in particular, the non-Gaussian parameter $f_{NL}^{\rm local}$ 
can be large if $\lambda$ is very small.
Note, however, that $\lambda$ cannot be too small.
For $\sigma\gg H$ which is necessary for the field
$\chi$ to work as a water-fall field, we must have 
$\lambda^{1/2}\gg10^{-6}$ from Eqs.~(\ref{HandL}), hence
$\lambda^{1/4}\gg10^{-3}$. 

\section{conclusion}
\label{sec:conclusion}

We have investigated analytically the curvature perturbation from a model
of multi-field hybrid inflation, which we call ``{\it multi-brid inflation}''.

First we have considered a general condition for a model
to be exactly soluble under the slow-roll approximation.
Then we have presented a multi-brid inflation model
in which the curvature perturbation is analytically
computable to full nonlinear order. We have described a method to
calculate the curvature perturbation analytically.
Here we have noted that the coupling of the (multi-component)
inflaton field to a water-fall field that terminates inflation
plays a significant role in the determination of
the curvature perturbation.

Then as a specific example, we have focused on
a two-brid inflation model, and derived an analytical expression
for the curvature perturbation to second order in the field
fluctuations expressed in terms of the model parameters.
Using this expression, we have given analytical formulas
for the spectrum of the curvature perturbation ${\cal P}_S$,
the spectral index $n_S$, the tensor-to-scalar ratio $r$,
and the non-Gaussian parameter $f_{NL}^{\rm local}$.
We have noted that adding only one extra component to
the inflaton field increases substantially the degrees
of freedom in the output parameters that are observationally
constrained.
We have found that our two-brid model can explain the
recent WMAP observations~\cite{WMAP3y,Dunkley:2008ie,Komatsu:2008hk}.

One possibly subtle issue we have not discussed in this paper is 
the effect of isocurvature perturbations at or after reheating.
When the universe is reheated by the water-fall field (and
perhaps by the inflaton field as well),
the abundance of created particles may depend on the initial values
of the components of the inflaton field. This can be regarded as a
two-field version of the modulated
fluctuations~\cite{Dvali:2003em,Kofman:2003nx}.
This would lead to additional power to the curvature perturbation
spectrum and probably to additional non-Gaussianity.
This is an interesting effect which deserves further study.
But from the point of view of the model we discussed in this
paper, this dependence must be weak enough for our model to be viable.
This effect must be carefully examined when we actually attempt to carry 
out the model-construction.
In fact, Barnaby and Cline found that the effect is
significant and leads to non-trivial constraints
on some models of hybrid inflation~\cite{BarnabyCline}.

Recently a lot of efforts have been paid in the construction of 
stringy inflation models~(see e.g.~\cite{Bond:2006nc,Haack:2008yb}
and references therein; for reviews on stringy cosmology,
see e.g.~\cite{stringcos}). Many of these models fall into a class
called brane inflation~\cite{Dvali:1998pa}, in which a brane is
attracted to another brane (or anti-brane) in higher dimensions
and the distance between the branes play the role of the inflaton.
In these models, inflation ends with a collision of the branes. 
Then the collided branes annihilate to heat up the universe.
Thus, in the 4-dimensional effective picture, brane inflation is
the same as hybrid inflation. Therefore it seems quite possible to
construct a stringy model that gives rise to multi-brid inflation.
This is currently under investigation.

\acknowledgements
I am extremely grateful to R. Kallosh and A. Linde for 
fruitful discussions and for their invaluable comments and
suggestions. I am also grateful to D. Lyth for very useful
comments.
This work was supported by the Yukawa International Program for
Quark-Hadron Sciences,
by JSPS Grant-in-Aid for Scientific Research (B) No.~17340075, 
and (A) No.~18204024, and by JSPS Grant-in-Aid for 
Creative Scientific Research No.~19GS0219.

\appendix
\section{Calculation of $\bm{\delta N}$ to second order}
\label{sec:detail}

Let us first evaluate the perturbation in $\gamma$ to second order.
{}Setting $\delta\gamma=\delta_1\gamma+\delta_2\gamma$, where
$\delta_1\gamma$ and $\delta_2\gamma$ are of linear and
second orders, respectively, we take
the perturbation of Eq.~(\ref{constofmotion}) to second order,
assuming $\delta\phi_1$ and $\delta\phi_2$ are of linear order.
We have
\begin{eqnarray}
\frac{\delta\phi_1}{m_1}-\frac{\delta\phi_2}{m_2}
=-\sigma\left(\frac{\sin\gamma}{g_1m_1}+\frac{\cos\gamma}{g_2m_2}\right)
(\delta_1\gamma+\delta_2\gamma)
-\frac{\sigma}{2}\left(\frac{\cos\gamma}{g_1m_1}-\frac{\sin\gamma}{g_2m_2}\right)
(\delta_1\gamma)^2\,.
\label{dgammaeq}
\end{eqnarray}
The linear part of the above equation determines $\delta_1\gamma$.
We find
\begin{eqnarray}
\delta_1\gamma=-\frac{g_1g_2}{\sigma}
\frac{m_2\delta\phi_1-m_1\delta\phi_2}{g_1m_1\cos\gamma+g_2m_2\sin\gamma}\,.
\label{d1gamma}
\end{eqnarray}
Then collecting the second order terms in Eq.~(\ref{dgammaeq}),
we find
\begin{eqnarray}
\delta_2\gamma
&=&\frac{1}{2}
\frac{g_1m_1\sin\gamma-g_2m_2\cos\gamma}{g_1m_1\cos\gamma+g_2m_2\sin\gamma}
(\delta_1\gamma)^2
\cr\cr
&=&\frac{g_1^2g_2^2}{2\sigma^2}
\frac{g_1m_1\sin\gamma-g_2m_2\cos\gamma}{(g_1m_1\cos\gamma+g_2m_2\sin\gamma)^3}
(m_2\delta\phi_1-m_1\delta\phi_2)^2\,.
\label{d2gamma}
\end{eqnarray}

Now it is straightforward to calculate $\delta N$. 
Expanding eq.~(\ref{N2brid}) to second order in the perturbation,
and substituting Eqs.~(\ref{d1gamma}) and (\ref{d2gamma}) into the resulting
equation, we obtain
\begin{eqnarray}
\delta N
=\frac{\delta\phi_1\,g_1\cos\gamma+\delta\phi_2\,g_2\sin\gamma}
{m_1g_1\cos\gamma+m_2g_2\sin\gamma}
+\frac{g_1^2g_2^2}{2\sigma}
\frac{(m_2\delta\phi_1-m_1\delta\phi_2)^2}
{(m_1g_1\cos\gamma+m_2g_2\sin\gamma)^3}\,.
\label{dN2ndapp}
\end{eqnarray}
This is recapitulated in Eq.~(\ref{dN2nd}).

It may be noted that as long as explicit computations of $\delta N$
are concerned, it is easier to go back to the original slow-roll
equations (\ref{mbrideom}). Integrating them are trivial.
We immediately find
\begin{eqnarray}
\frac{\phi_1-\phi_{1,f}}{m_1}=N\,,
\quad
\frac{\phi_2-\phi_{2,f}}{m_1}=N\,,
\label{phiAsol}
\end{eqnarray}
where 
$\phi_{1,f}$ and $\phi_{2,f}$ are parameterized in terms of $\gamma$ as
in Eq.~(\ref{gammadef}),
\begin{eqnarray}
\phi_{1,f}=\frac{\sigma}{g_1}\cos\gamma\,,
\quad
\phi_{2,f}=\frac{\sigma}{g_2}\sin\gamma\,.
\label{phif}
\end{eqnarray}
One can then take the perturbation of either the first or the second
equation in Eq.~(\ref{phiAsol}) to second order, with $\delta\phi_{1,f}$
and $\delta\phi_{2,f}$ given by
\begin{eqnarray}
\delta\phi_{1,f}
&=&\frac{\sigma}{g_1}
\left[-\sin\gamma\,\left(\delta_1\gamma+\delta_2\gamma\right)
-\cos\gamma\frac{(\delta_1\gamma)^2}{2}\right]\,,
\cr\cr
\delta\phi_{2,f}
&=&\frac{\sigma}{g_1}
\left[\cos\gamma\,\left(\delta_1\gamma+\delta_2\gamma\right)
-\sin\gamma\frac{(\delta_1\gamma)^2}{2}\right]\,.
\label{dphif}
\end{eqnarray}
The result is the same whatever method one adopts.
In fact this may be used to check the calculation.
For example, taking the perturbation of the first equation in (\ref{phiAsol})
and using the first line of Eq.~(\ref{dphif}),
one finds
\begin{eqnarray}
\delta N=\frac{\delta\phi_1}{m_1}-\frac{\delta\phi_{1,f}}{m_1}
=\frac{\delta\phi_1}{m_1}+\frac{\sigma}{m_1g_1}\sin\gamma\,\delta_1\gamma
+\frac{\sigma}{m_1g_1}
\left[\sin\gamma\,\delta_2\gamma+\cos\gamma\frac{(\delta_1\gamma)^2}{2}\right]\,.
\end{eqnarray}
Inserting Eqs.~(\ref{d1gamma}) and (\ref{d2gamma}) into this equation,
one can check that the result agrees with Eq.~(\ref{dN2ndapp}).

Now let us turn to the evaluation of the correction term after inflation,
$\delta N_c$. Taking the perturbation of $N_c$ in Eq.~(\ref{Ncorrect}),
we find
\begin{eqnarray}
\delta N_c&=&\frac{1}{4}
\left[m_1\delta\phi_{1,f}+m_2\delta\phi_{2,f}\right]
=-\frac{\sigma}{4}
\left(\frac{m_1}{g_1}\sin\gamma-\frac{m_2}{g_2}\cos\gamma\right)\delta_1\gamma
\cr\cr
&&-\frac{\sigma}{4}
\left[
\left(\frac{m_1}{g_1}\sin\gamma-\frac{m_2}{g_2}\cos\gamma\right)\delta_2\gamma
+\left(\frac{m_1}{g_1}\cos\gamma+\frac{m_2}{g_2}\sin\gamma\right)
\frac{(\delta_1\gamma)^2}{2}
\right].
\end{eqnarray}
Inserting $\delta_1\gamma$ and $\delta_2\gamma$ given by Eqs.~(\ref{d1gamma})
and (\ref{d2gamma}) into the above, we obtain
\begin{eqnarray}
\delta N_c
=\frac{1}{4}
\frac{m_1g_2\sin\gamma-m_2g_1\cos\gamma}{m_1g_1\cos\gamma+m_2g_2\sin\gamma}
(m_2\delta\phi_1-m_1\delta\phi_2)
-\frac{g_1^2g_2^2}{8\sigma}
\frac{(m_1^2+m_2^2)(m_2\delta\phi_1-m_1\delta\phi_2)^2}
{(m_1g_1\cos\gamma+m_2g_2\sin\gamma)^3}\,.
\end{eqnarray}
Comparing this with Eq.~(\ref{dN2ndapp}), we see that the contribution
of $\delta N_c$ to $\delta N$ is suppressed by a factor quadratic in
$m_1$ and/or $m_2$. Hence, under the assumption that $m_1$ and $m_2$ are
sufficiently smaller than unity, which is necessary for the slow-roll
condition to hold, we may neglect the correction term $\delta N_c$.

Finally, let us evaluate $\delta N$ up to a surface of constant potential
$V=\mbox{const.}=V_*$ during inflation, which we denote by $\delta N_*$,
and identify the contribution from the end of inflation, which we
denote by $\delta N_e$.
On superhorizon scales, and uder the slow-roll approximation,
$\delta N_*$ is equal to the curvature perturbation on the uniform
density slice, which is usually denoted by $\zeta$.

The slice $V=V_*$ is given by
\begin{eqnarray}
m_1\phi_{1,*}+m_2\phi_{2,*}=\ln(V_*/V_0)\equiv C\,.
\end{eqnarray}
One can parametrize $\phi_{A,*}$ ($A=1,2$) with a parameter $t$ as
\begin{eqnarray}
\phi_{1,*}=\frac{1}{2m_1}(C+2t)\,,
\quad
\phi_{2,*}=\frac{1}{2m_2}(C-2t)\,.
\end{eqnarray}
This gives
\begin{eqnarray}
\delta\phi_{1,*}=\frac{1}{m_1}\delta t\,,
\quad
\delta\phi_{2,*}=-\frac{1}{m_2}\delta t\,.
\end{eqnarray}
Thus taking the perturbation of the solutions (\ref{phiAsol}),
with $\phi_{A,f}$ replaced by $\phi_{A,*}$ and $N$ by $N_*$,
we obtain two expressions for $\delta N_*$,
\begin{eqnarray}
\delta N_*
&=&\frac{\delta\phi_1}{m_1}-\frac{\delta\phi_{1,*}}{m_1}
=\frac{\delta\phi_1}{m_1}-\frac{\delta t}{m_1^2}\,,
\cr\cr
\delta N_*
&=&\frac{\delta\phi_2}{m_2}-\frac{\delta\phi_{2,*}}{m_2}
=\frac{\delta\phi_2}{m_2}+\frac{\delta t}{m_2^2}\,.
\label{deltaN2}
\end{eqnarray}
{}From these equations, we find
\begin{eqnarray}
\delta t=\frac{m_1m_2}{m_1^2+m_2^2}(m_2\delta\phi_1-m_1\delta\phi_2)\,.
\end{eqnarray}
Inserting this to either one of Eqs.~(\ref{deltaN2}),
we obtain
\begin{eqnarray}
\delta N_*=\frac{m_1\delta\phi_1+m_2\delta\phi_2}{m_1^2+m_2^2}\,.
\label{dNstar}
\end{eqnarray}
As clear from this expression, in our model,
there is no non-Gaussianity from the evolution during inflation.
This is in agreement with the discussion based on the equations in the
$q_A$ space given near the end of Section~\ref{sec:model}.

Subtracting $\delta N_*$ from $\delta N$ given in Eq.~(\ref{dN2ndapp}),
we obtain the contribution from the end of inflation $\delta N_e$.
The result is
\begin{eqnarray}
\delta N_e&=&\delta N-\delta N_*
\cr
&=&\frac{(m_2g_1\cos\gamma-m_1g_2\sin\gamma)(m_2\delta\phi_1-m_1\delta\phi_2)}
{(m_1^2+m_2^2)(m_1g_1\cos\gamma+m_2g_2\sin\gamma)}
+\frac{g_1^2g_2^2}{2\sigma}
\frac{(m_2\delta\phi_1-m_1\delta\phi_2)^2}
{(m_1g_1\cos\gamma+m_2g_2\sin\gamma)^3}\,.
\end{eqnarray}


\end{document}